\documentstyle[12pt,aaspp4]{article}

\def\ltsima{$\; \buildrel < \over \sim \;$}
\def\simlt{\lower.5ex\hbox{\ltsima}}	
\def\gtsima{$\; \buildrel > \over \sim \;$}
\def\simgt{\lower.5ex\hbox{\gtsima}}	

\def\ref{\par\noindent\hangindent 20 pt}

\def\mincir{\ \raise -2.truept\hbox{\rlap{\hbox{$\sim$}}\raise5.truept 
\hbox{$<$}\ }}	%
\def\magcir{\ \raise -2.truept\hbox{\rlap{\hbox{$\sim$}}\raise5.truept %

\hbox{$>$}\ }}
\def\ea {et al. }
\def\cosmo{H$_0$ = 50~km~s$^{-1}$~kpc$^{-1}$ and q$_0 = 0$} 
\def\asec{$^{\prime\prime}$}	
\def\asecd{$^{\prime\prime}\!\!$}	

\received{}
\accepted{}
\journalid{}{}
\articleid{}{}

\input{psfig.tex}

\begin{document}

\title{\bf The HST Survey of BL~Lacertae Objects. III. Morphological 
Properties of Low-Redshift Host Galaxies}

\author{Renato Falomo} 
\authoraddr{Osservatorio Astronomico di Padova, Vicolo dell'Osservatorio 5, 
35122 Padova, Italy}
\affil{Osservatorio Astronomico di Padova}

\author{Riccardo Scarpa\altaffilmark{1}} 
\authoraddr{Space Telescope Science Institute, 3700 San Martin Dr., 
Baltimore, MD 21218, USA}
\affil{Space Telescope Science Institute}
\altaffiltext{1}{also at Department of Astronomy, Padova University, Vicolo 
dell'Osservatorio 5, 35122 Padova, Italy}

\author{Aldo Treves} 
\authoraddr{Universit\`a dell'Insubria, via Lucini 3 Como, Italy}
\affil{Universit\`a dell'Insubria, Como, Italy}

\author{C. Megan Urry} 
\authoraddr{Space Telescope Science Institute, 3700 San Martin Dr., Baltimore,
MD 21218, USA}
\affil{Space Telescope Science Institute}


\begin{abstract}
 
We report on the optical properties of a sample of 30 BL Lac host
galaxies in the redshift range $0.03<z<0.2$, as derived from HST
observations. All galaxies are fully resolved in the WFPC2 (F702W
filter) images, allowing a quantitative analysis in two
dimensions. Most and possibly all these galaxies have characteristics
very similar to those of ``normal'' giant ellipticals. The luminosity,
ellipticity, isophote twisting and amount of disky or boxy isophotes
are consistent with those found in non-active ellipticals and in radio
galaxies. In all cases
the BL Lac nucleus is well centered in the main body of its host
galaxy, a result that argues strongly against the microlensing
hypothesis for any significant fraction of the population. A search
for faint sub-structures in the host galaxies has not revealed
notable  signatures of tidal distortions or sub-components (faint
disks, bars, X features, etc.), and with only one exception, there are
no prominent dusty features in the central regions. Instead, the BL
Lac host galaxies are smooth and unperturbed, suggesting that strong external
gravitational interactions are not important to ongoing activity. A
careful examination of the environment around the nucleus, however, shows a
high incidence of close companion objects, whose nature remains
unclear pending spectroscopic observations.

\end{abstract}

{\underline{\em Subject Headings:}} 
BL~Lacertae objects:general --- BL~Lacertae objects: host galaxies
Galaxies: structure --- Galaxies: interactions

\section{Introduction}

The Hubble Space Telescope (HST) snapshot imaging survey of BL Lac
objects\footnote{Based on observations made with the NASA/ESA Hubble
Space Telescope, obtained at the Space Telescope Science Institute,
which is operated by the Association of Universities for Research in
Astronomy, Inc., under NASA contract NAS~5-26555.} has produced a
large, homogeneous data set of over one hundred high-resolution 
images for the study of the environments of BL Lac nuclei. The
targets were drawn from various samples and include sources in a wide
range of redshift ($z$ from 0.03 to $\sim$ 1.4) as well as objects
with unknown $z$. Results from the survey are given in Scarpa et
al. (2000; hereafter, Paper I), and Urry \ea (2000; hereafter, Paper II).
Previous HST imaging of BL Lacs had been done only for a small number of
sources (see Falomo \ea 1997; Jannuzi \ea 1997; Urry \ea 1999) but
demonstrated the capabilities of HST for studying the environs of BL
Lacs.

The HST snapshot images have two main advantages with respect to
ground-based images (e.g., Falomo 1996; Wurtz, Stocke, \& Yee 1996;
Falomo \& Kotilainen 1999): an order of magnitude improved spatial
resolution and very good homogeneity of data. On the other hand the
snapshot images were obtained with relatively short exposure times
(typically minutes) and therefore are not very deep. This sets limits
to the detection of very faint extended features, a well-known
problem for HST images (see Falomo \& Kotilainen 1999 for a direct comparison of HST and ground-based images of BL Lacs).

Our first papers on the overall properties of the BL Lac objects derived
from HST data focused on the determination of
the absolute magnitude, scale-length, and type of the host
galaxies, as well as the luminosity of the nuclear source (see Paper II). This was
based on accurate decomposition of the azimuthally averaged luminosity
radial profiles which allowed the study of the sources down to surface
brightness $\mu_R \approx $25~mag~arcsec$^{-2}$. Additional
morphological properties of the surrounding nebulosity, such as
ellipticity, isophotal twisting, isophote shape (disky versus boxy),
presence of dust, etc., can only be derived from a full
two-dimensional approach, which is possible for a subset of bright
nearby host galaxies.

For elliptical galaxies it has been shown that isophote shapes are
related to internal kinematics and to the radio and X-ray emission
properties (e.g. Bender, D\"obereiner, \& M\"ollenhoff 1987). 
Moreover significant distortions of the isophotes and/or
twisting may be produced by strong galaxy interactions and/or merger 
events. It is therefore of interest to investigate these observable
features in the galaxies hosting BL Lac nuclei in order to understand
the (possible) role of galaxy structure on the nuclear activity.
This kind of analysis has been performed so far only for a small
number of nearby BL Lac objects (e.g., Falomo 1991,1996; Heidt et al. 
1999) because the presence of the bright nucleus often hinders 
a fully detailed morphological study.

In this paper we present the results of a two-dimensional analysis for 30 nearby
($z \lesssim 0.2$) snapshot BL Lacs, i.e., those with the highest
signal-to-noise ratios and the largest extent (see Table~1). For 25 of
the 110 objects observed in the HST snapshot survey (Paper I)
the redshift is unknown. Of these 17 are spatially
unresolved and are therefore most probably at $z > 0.2$. For the
remaining 8 resolved objects with unknown redshift, based on the
apparent magnitudes of the nebulosities (see Paper I) and
assuming the luminosity of the host is consistent with that of other snap BL
Lacs (see Fig. 1), one expects that at most a few can be at $z < 0.2$.
Therefore the objects discussed here represent practically the totality of
$z < 0.2$ BL Lacs from our original sample of 110 objects.

For these low-$z$ objects it is possible with HST to investigate
features and structures in the host galaxy that are undetectable with 
ground-based observations. The host galaxy can be investigated down to less
than 1~kpc from the nucleus (corresponding to 0\asecd.3 at $z = 0.2$).
In particular one can search for slight off-centering of the nucleus
with respect to the galaxy, determine the presence of sub-components
in the host galaxy, investigate the detailed shape of the isophotes, 
and search for dusty features and close companions.

Because the two broad components in BL~Lac spectral energy
distributions (SEDs) have peak power outputs ($\nu L_\nu$) at
wavelengths that increase systematically with luminosity (e.g.,
Ulrich, Maraschi \& Urry 1997), BL~Lacs found in radio and X-ray
surveys have generally different SEDs.  Radio-selected objects tend to
have SEDs peaking at infrared-optical wavelengths and in the MeV-GeV
gamma-ray band (low-frequency-peaked BL~Lacs, or LBL) and exhibit
luminosities approaching those of quasars.  In contrast,
X-ray-selected BL Lacs have SEDs peaking at UV-X-ray wavelengths and
again at TeV energies (high-frequency-peaked BL~Lacs, or HBL) and are
generally less luminous.  Of the 30 sources in the present study, 6
are of LBL type , the rest being HBL (Padovani \& Giommi 1995).  In
the following analysis we have not addressed whether morphological
properties are different for the two types of BL Lacs, having found
from our earlier analysis that the overall host galaxy characteristics
do not depend on type (Paper II).

In \S~2 we briefly describe the observations and our data
analysis. \S~3 gives the results on the host galaxies properties and
shows the comparison with non-active ellipticals and radio galaxies.
Finally in \S~4 we summarize the results and draw the main conclusions
from this study. Throughout the paper we used \cosmo.

\section{Observations and Data Analysis}

The snapshot observations of BL Lacs have been described in detail in
Paper I, therefore only the main steps are reviewed here.  All objects considered here 
were observed with the WFPC2 camera of the {\it HST} in the F702W
filter. Each final image was obtained from the combination of
typically three exposures for a total integration time of about 5-20
minutes (see Table 1 in Paper I for details).  The Point Spread
Function (PSF) used is a combination of a model derived from Tiny Tim
package (Krist 1995) plus a diffuse exponential tail that accounts for
the first order scattered light (see Urry \ea 1999 and Paper I for
details). Photometric calibration was taken from Holtzman \ea (1995).

We performed a two-dimensional surface photometry analysis using an
interactive numerical mapping package (AIAP; Fasano 1994), which
produces an isophotal map of the image. This map was masked in order
to avoid regions contaminated by companion objects, diffraction
spikes, and any other extraneous features visible in the images. The
isophotes were then fitted with ellipses down to $ \mu_{R}$ =
22--23~mag~arcsec$^{-2}$ (see Fig.~2 for an example). There are five
geometric free parameters per isophote: semi-major axis ($a$), center
position, ellipticity ($\epsilon$), and position angle (PA), which
allow us to characterize the morphological and the photometric
properties of the galaxy.

From this analysis we then derived the average position angles and
ellipticity as a function of the generalized radius, $r =
a\times(1-\epsilon)^{1/2} $, and investigated their dependence on $r$
(isophote twisting and ellipticity profiles). In addition we examined
the possible displacement of the centers of isophotes and the
centering of the nucleus with respect to the galaxy.

\section{Results}

\subsection{Luminosity Profiles and Two Dimensional Modeling}

For each object we derived a luminosity profile from the
two-dimensional isophotal analysis and compared with that derived from
the azimuthal average reported in Paper I. This was possible down to $
\mu_{R} \approx$ 22--23~mag~arcsec$^{-2}$ since the fainter outer
regions are too noisy to be investigated with the isophotal map. For
all sources we find that the luminosity profile derived from the fit
with ellipses of the isophotes is always in excellent agreement with
that extracted from the azimuthal average (see example in Fig.~3).

We also fitted the isophotes with a two-dimensional galaxy model
(either a de~Vaucouleurs law or an exponential disk) convolved with
the PSF plus a point source modeled by a scaled PSF. In all cases the
de Vaucouleurs law gives a better fit to the data than the exponential
disk model, in agreement with Papers I and II. This result is
supported by the lack of spiral structures seen in the WFPC2 images
and by the {\it diskiness} of the isophotes (see discussion in the
following paragraphs). Absolute magnitudes and scale lengths of the
host galaxies derived from the two-dimensional fitting are also in
excellent agreement with the results of one-dimensional profile
fitting (Paper I), and are consistent with the results from our deep
HST images for a small number of BL Lac objects (Falomo \ea 1997).

\subsection{Ellipticity }

Since the ellipticity $\epsilon$ is in general dependent on the
distance from the galaxy center, we report its value at the effective
radius. However, because of the relatively high surface brightness
limit, in some objects the ellipticity at the effective radius is not
measurable and in these cases we took the maximum ellipticity. 

The values of $\epsilon$ are reported in Table~1. The
error on observed $\epsilon$ for isophotes far from the nuclear
regions is typically 0.02 (see Fasano \& Bonoli 1990 for discussion).
We show in Figure~4
the distribution of ellipticity for our 30 low-redshift BL Lacs compared
with that of 200 nearby normal (radio-quiet) ellipticals (Fasano \& Vio
1991) and of a sample of 79 low-redshift radio galaxies (Govoni \ea
2000). Both normal ellipticals and radio galaxies exhibit the same
ellipticity distribution: $<\epsilon>_{Ell}$= 0.22 $\pm$ 0.13 and
$<\epsilon>_{RG}$= 0.21 $\pm$ 0.12. The average ellipticity for BL Lac
hosts, $<\epsilon>_{BLL}$ = 0.16 $\pm$ 0.09, is slightly smaller than
that of non-radio ellipticals and radio galaxies but is still consistent
with being drawn from the same population ($P_{KS} \sim 0.1$).

\subsection{Isophotal Displacement and Nuclear De-Centering}

Displacement of isophotes with respect to the center may represent
either a global distortion of the host galaxy, due to recent tidal
interactions with close companions (e.g., Aguilar \& White 1986), or a
consequence of gravitational microlensing (Ostriker and Vietri
1985). In the first case one would expect non-concentric isophotes to
exhibit an asymmetry of distribution. To quantify this asymmetry we
have computed the dimensionless parameter $\delta \equiv
\sqrt{(X_c-X_o)^2+(Y_c-Y_o)^2}/r$, where $X_c, Y_c$ are the centers of
isophotal ellipses, $X_o, Y_o$ is the location of the nucleus, and $r$
is the radial distance to the particular elliptical isophote (defined
earlier). Since $\delta$ may slightly change with $r$ we took the
value at the effective radius R$_e$ of the galaxy or, if R$_e$ is
beyond our measurements, we took the average value excluding the
values at $r < 1$ arcsec.

In Figure~5 we show the distribution of $\delta$ derived from our
sample of low-redshift BL Lacs compared with that of radio galaxies
from Govoni et al. (2000). The average value for our sample is:
$<\delta>_{BLL} = 0.027 \pm 0.01$ (median $\delta$ is 0.030),
marginally consistent with zero but suggestive of small distortions in
the galaxy isophotes.  This is similar to the value reported for low
$z$ radio galaxies, $<\delta>_{RG} = 0.03 \pm 0.04$ (Govoni et
al. 2000), using a homogeneous procedure, while for 40 normal
ellipticals observed by Sparks et al. (1991) from the ground, Colina
\& de Juan (1995) give $<\delta>_{Ell}$ = 0.02 (see  in Govoni et
al.  discussion and comparison with other samples of radio
galaxies). This is consistent with the idea that galaxies hosting BL
Lac nuclei and/or radio sources have indistinguishable structures and
have undergone a similar degree of interaction.

In the microlensing scenario (Ostriker \& Vietri 1985, 1990), what
appears to be the host galaxy is actually an intervening galaxy whose
stars are microlensing a background quasar. Since the alignment will
not be perfect the galaxy will generally be off-center with respect to
the point source, typically by $\gtrsim 0.5$~arcsec (Merrifield
1992). To investigate this possibility we computed for each object the
limit of off-centering, $\Delta = \vert (x,y)_{PSF} - (x,y)_{gal}
\vert$, comparing the PSF location with that of the galaxy (derived by
averaging of isophotal centers after subtraction of a scaled nuclear
point source and exclusion of the central circle of radius 0.5
arcsec).  With HST images the off-centering can be computed with
unprecedented accuracy given the significantly better spatial
resolution. The centers of nucleus and the host galaxy have a typical
uncertainty of 0.2 and 0.4 pixels, respectively, so the net
uncertainty is typically $\sim 0.05$~arcsec.

None of the objects in our (low-redshift) sample show significant ($>
2 \sigma$) displacement of the point source with respect to the galaxy
(values of $\Delta$ are given in Table~1); the observed displacements
are always $< 0.1$~arcsec and therefore consistent with zero
off-centering. Since the low-redshift BL Lacs were the prime
candidates for microlensing according to Ostriker \& Vietri (1985),
our result strongly rules out the microlensing hypothesis for BL Lac
objects in general.

\subsection{Isophotal Twisting}

For each host galaxy having ellipticity larger than 0.15 we have
estimated the isophote twisting, $\Delta\theta$, computing the
difference of position angles over the whole range of observed surface
brightness (see Table 1). For the others the small ellipticity
introduces large errors on the position angle so that it is not
possible to get a reliable measurement of the isophotal twist.

The largest observed twist is $\sim 30^\circ$ in 1853+671; this
value is however uncertain because of the relatively small ellipticity
and short exposure time. We found five other objects (corresponding to
$\sim$20\% of the sample) with isophote twists larger than 15$^\circ$.
This is consistent (see Fig. 6) with what was found for a sample of 43 isolated
(normal) ellipticals (Fasano \& Bonoli 1989) and for 79 low-redshift
radio galaxies (Govoni et al. 2000) using the same method to derive
$\Delta$PA.

A small amount of isophote twisting may be explained simply by the
triaxiality of the galaxies, while larger twists may be due to the
effects of tidal interaction with companion galaxies (Kormendy 1982). 
The lack of large twists in our sample is consistent with the
overall smoothness (unperturbed shape) of the host galaxies.

\subsection{Isophotal Shapes: Disky and Boxy}

To check for small deviations from purely elliptical isophotes we
analyzed the amplitude of the fourth cosine component, $C_4$, of the
Fourier fit to the isophotes (e.g., Bender \& Saglia 1998). A
significant positive value of $C_4$ corresponds to {\it disky}-shaped
isophotes while a negative value indicates a {\it boxy}-shaped
structure. 

We found that 80\% of the sources show a $C_4$ amplitude (at the
effective radius) smaller than 1\% while no object exhibits $\vert C_4 \vert$
larger than 3\%. The distribution of $C_4$, shown in Figure~7, is
symmetric around zero (mean value $<C_4>_{BLL} = 0.03$\%, rms dispersion 0.7\%)
and since the size of the error is around 0.5\% we conclude that in
these galaxies there is no systematic deviation of isophotal shape
from a simple ellipse. Similar distributions are observed in a sample
of low-redshift radio galaxies using the same method 
($<C_4>_{RG} = 0.04$\% $\pm 1.2$\%; Govoni et al. 2000) 
and also for radio galaxies in
rich clusters (Ledlow and Owen 1995), as well as for normal
ellipticals (Jorgensen et al. 1995).

\subsection{Structures in the Host Galaxies}

A first look at the raw images of BL Lacs as well as the surface
photometry analysis above indicates that galaxies hosting BL Lacs are
rather smooth and unperturbed (see images in Paper I).  However, faint
sub-structures could be hidden by the smooth contribution of the
stellar component, so we subtracted a two-dimensional model of the
galaxy using the best-fit parameters from the surface photometry
analysis. This procedure enhances faint structures such as companions,
jet-like features, or any other high-contrast feature superposed onto
the image of the galaxy (e.g., Faundez-Abans \& Oliveira-Abans 1998,
and references therein). Clearly any region masked out during
isophotal analysis (including the diffraction pattern of PSF spikes)
will come out from this procedure. Moreover due to the sharpness of
the nucleus and the insufficient sampling of the PC it was not
possible to model the PSF adequately with this technique and some
residuals (typically diffraction spikes) remain close to the
nucleus. This has very little impact on our conclusions since
structures very close to the nucleus are not considered.

In Figure~8 we show three examples of the images of the BL Lacs before
(left) and after (right) subtraction of the galaxy model. Panel (a)
shows the optical jet of PKS 0521-36 after subtraction of the
galaxy-plus-point-source model.  Panel (b) shows the image of BL Lac
itself (2200+420) which, after subtraction of the model, does not
exhibit significant features. Finally panel (c) shows an example
(H~2356-309) of a companion at $\sim 1$~arcsec south-west of the
nucleus. In all three subtracted images the emission from the
diffraction spikes (not modeled) is clearly visible.

The BL Lac objects in the present sample
span the redshift range $0.03 \lesssim z \lesssim 0.2$.
The usable field of view of the PC camera corresponds thus to
projected linear size of $\sim 25$ to $\sim 135$~kpc. At HST
resolution therefore we are practically able to explore structures in
the host galaxies as close as 0.2 -- 0.9~kpc (depending on $z$) to the
nucleus (we assume that within 5 pixels from the center the signal is
strongly dominated by the nuclear point source).

In three cases (0521-36, 3C 371 and 2201+044) an optical jet is clearly
visible (see Scarpa \ea 1999b for a detailed study of these objects).
One object (0806+524) has arc-like emission at $\sim 2$~arcsec from
the nucleus, which could be a shell around the galaxy, a feature not
uncommon among ellipticals (see Scarpa \ea 1999a for details). No other
faint structures (shells, bars or X features) are detected in the
observed sources.

HST observations of nearby radio galaxies have shown that dusty features
are present in the central regions of most of the observed objects
(Kleijn et al. 2000). These features often take the form of disks or
lanes but sometimes the distribution is irregular. The observed dust
pattern is generally confined within a region typically less
than 1.5~kpc (in most cases the physical sizes of disks and lanes range
from 0.3 to 1.5~kpc). Since BL Lacs are believed to be radio galaxies
with the jet pointing toward us, we should observe similar dusty
features also in our objects. However, two difficulties make this
detection less likely in our sample objects. First, the bright nucleus
tends to out-shine the central regions of the host galaxy. Second, the
average distance of our objects is larger than that of comparison
samples of radio galaxies observed with HST, making the angular size of
dusty features smaller. At $z = 0.05$, 1~kpc corresponds to 
$\sim 0.75$~arcsec; at this distance from the center the light from the
nucleus is normally fainter than that from the host galaxy (see Fig.~1
in Paper I) so that these features could in principle be observable at
the very lowest redshift but become progressively more difficult to 
detect at higher $z$. 

In order to explore this possibility we have therefore considered only
the 9 objects at $z < 0.1$ and searched for dusty features in the
central regions after subtraction of the bright point source from the
original image.  In only one object (1959+650), which is at redshift
$z = 0.048$, is a significant dust lane observed at $\sim0.8$~arcsec
(corresponding to $\sim 1$~kpc) north of the nucleus and roughly
aligned with the major axis of the galaxy. This feature was also noted
by Heidt et al.  (1999) on ground-based images and is discussed in our
previous work (Scarpa et al. 1999a) on peculiar objects observed in
the snapshot survey. For the rest of the eight $z < 0.1$ sources no
clear signature of dust is found. Note that four of these objects are
at a redshift similar to or smaller than that of 1959+650. This lack
of detection suggests that the relevance of dusty features may be
different in BL Lacs and radio galaxies, with the caveat of the small
statistics considered.  Moreover, note that looking at features close
to the nucleus requires not only high spatial resolution but also a
PSF with sufficient sampling and faint wings. HST with the new
Advanced Camera for Surveys High Resolution Camera offers two-times
better sampling, and has a coronograph that depresses the PSF wings by
factors of a few (and obscures the central 0.9~arcsec), so it could
contribute to elucidating this point.

\section{Close Companions}

In a number of objects we noted the presence of faint companions
around the target, a feature that seems also common to the HST images
of quasars. We have systematically explored the near environments of
the 30 low-redshift BL Lac sources with the aim of identifying
potential companion objects.  For three cases (0706+592, 1440+122 and
2356-309) a very close ($\Delta <$ 1\asecd.2) compact companion is
detected. These companions are located at projected distances of the
order of 1--5~kpc from the nucleus. For five other BL Lac objects (see
Table~2) diffuse or compact companions are found within a projected
distance of $\lesssim 20$~kpc (if they are at the same redshift as the
BL Lac).

In order to evaluate the statistical significance of the occurrence of
these close companions we derived the expected number of objects
within a given radius from the BL Lac based on the number counts
observed in our PC frames well away from the BL Lac.  The observed
average density of objects with $19.5< m_R < 23.5$~mag is $9 \pm
5$~arcmin$^{-2}$, in good agreement with the observed galaxy density
in several high galactic latitude fields (Metcalfe et al. 1991).

At these faint magnitudes the counts at high galactic latitudes are
dominated by galaxies. The average galactic latitude of our fields is
$|b| = 45^{\deg}$ but two fields (2200+420, 2344+514) are at low
galactic latitude ($|b| \sim 10^{\deg}$) and one (1514--241) is at
$|b| \sim 27^{\deg}$ but close to the direction toward the galactic
center. The observed average density of stellar (unresolved) objects 
including all PC frames
is $3.2 \pm 4.2$~arcmin$^{-2}$, while excluding the three fields
mentioned above yields a star density of $2.1 \pm 2.6$~arcmin$^{-2}$. 
This is consistent with other estimates of the stellar density at high galactic
latitude (Hintzen, Romanishin \& Valdes 1991; Metcalfe \ea 1991).

Assuming the observed average density of faint ($19.5< m_R <
23.5$~mag) objects, 9~arcmin$^{-2}$, around 30 targets we would expect
to find 0.06 companions at $r < $0\asecd.5, while one (a compact
companion at 0.3~arcsec from 1440+122) is observed (Poisson
probability P=0.06). Within 1\asecd.5 we would expect to observe 0.6
objects while 3 are detected (P=0.02). At larger radii ($r<$5\asec) we
observe 11 companions when 6 are expected (P=0.02).  To test the
reality of this apparent excess, we counted the number of detected
objects within these same radii for random positions in each PC
frame. In those cases we found 0, 1, and 7 companions, respectively,
in agreement with the expected numbers of faint field objects.

Although the numbers are small and consequently the statistics are 
not very good, we conclude from our analysis 
that there is some indication that an association between BL Lacs 
and companions is not dominated by chance alignments. 

The identity or role of these companions is not yet clear, however.
In other BL Lac fields, optical spectroscopy of the companion has
sometimes proved the physical association with the active galaxy (Falomo 1996); in
other cases, they have proved to be stars or foreground or background
galaxies, fortuitously aligned with the BL Lac (e.g., Stickel et
al. 1991).  For most cases here, spectra of the companion objects are
not available and only statistical arguments suggest some are likely
associated with the BL Lac object.

Some of the close companions are resolved faint galaxies with absolute
magnitudes (assuming they are at the distance of the BL Lac object) in
the range $-21 < m_R < -19$~mag (Table~2).  This is consistent with
previous evidence that BL Lac objects inhabit environments with
higher-than-average galaxy densities, as in groups or poor clusters
(Pesce, Falomo \& Treves 1995; Wurtz et al. 1996).
In the two cases with more than one companion, 1229+645 and 1440+122,
the two companion objects are clearly different from one another, one
being resolved and the other unresolved,

We note that a significant number of the observed companions are
compact and appear point-like even at HST resolution.  Based on the
counts of Galactic stars, we would expect only 1 or 2 chance
alignments with stars while 6 compact companions are observed
(Table~2).  The excess of compact companions is in fact more
significant than the excess of resolved companions (galaxies),
especially since none of the compact sources belong to the low
galactic latitude fields where the star density is higher than the
average value.

The nature of these unresolved companions is unclear.  If they were at
the redshift of the BL Lac object, their magnitudes would be in the
range $-18 < m_R < -15$~mag, too luminous for globular clusters.
These magnitudes are consistent with dwarf ellipticals but the
observed sources are probably too compact, unless they contain weak
active nuclei (but they would have to be very weak indeed).  One can
ask whether some of these compact sources could be supernovae of type
Ia. The SN Ia rate in early type galaxies is estimated to be about
0.1--0.2 SN/100yr/10$^{10}$ L$_\odot^B$ (Cappellaro et al. 1999). This
means that in a massive galaxy of $M_R \sim$ -23.7 mag we expect to
have about 1-2 SN/100yr. Because the SN Ia at maximum reach $M_R
\approx$ -19.4~mag and our limiting magnitude for stellar objects is
$m_R \sim 24.5$~mag (it is $m_R \sim 24$~mag for extended objects), at
$z \sim 0.1$ we should be able to detect such SN for about 100 days.
Therefore we should find 0.2 SN in our 30 galaxies observed. Unless
the SN rate in these active galaxies is significantly higher than the
average for normal ellipticals, it is highly unlikely that the compact
companions are SN.

We believe the nature of all these companions, resolved or unresolved,
will remain uncertain until spectroscopy is performed.  Whatever their
identity, there is little evidence of tidal interaction in the BL Lac
host galaxy (appreciable isophote twisting, non-concentric isophotes,
etc.), which suggests the companions are not strongly perturbing it.
Gravitational interactions might have had a greater role in the past
in the formation and fueling of the active nucleus, and what we see
now could be a leftover of an earlier (close) interaction. One
interesting possibility suggested by numerical simulations (Bekki
1999) is that the close companions are the product of a past major
merger between gas-rich galaxies and also provide fuel for the nuclear
activity but are not necessarily linked to the formation of a massive
black hole in the nucleus, or to the sustenance of an active
relativistic jet.

\section{Summary and Conclusions}

We have presented detailed morphological analysis of HST images for 30
galaxies hosting BL Lac sources at redshift $z \lesssim 0.2$. The
galaxies were investigated with an order of magnitude better spatial
resolution than it is possible from the ground.

Our analysis indicates that in spite of the presence of the active
nucleus the host galaxy appears in most cases to be a completely ``normal''
elliptical.
Both the ellipticity and the isophotal shape distributions are similar
to those for radio galaxies and radio-quiet ellipticals. This
suggests that tidal interactions are very infrequent or are
short-lived with respect to the nuclear activity time scale.

We find no indication of displacement and/or off-centering of the galaxy
isophotes with respect to the nucleus, meaning the unresolved nuclear
source truly sits in the center of the galaxy. This rules out the
microlensing hypothesis for BL Lacs, which predicts frequent
off-centering of the nucleus. This does not exclude the possibility that
some objects could be lensed (see Scarpa et al. 1999a), but it cannot be
a widespread explanation of the BL Lac phenomenon.

An interesting result is the finding of excess close companions in several 
cases. A
high incidence of companions of BL Lacs was first suggested by (Falomo, 
Melnick \& Tanzi 1991) from sub-arcsec resolution ground-based imaging, and
was later confirmed by Pesce, Falomo and Treves (1995), Falomo (1996)
and Heidt et al. (1999). Close companions seem also to be common around
quasars imaged by HST (Bahcall \ea 1997; Disney \ea 1995). 

\acknowledgements

We thank G. Fasano for useful suggestions and support performing
isophotal (AIAP) analysis and E. Cappellaro and M. Turatto for
discussion about the SN rate. This work was partly supported by the
Italian Ministry for University and Research (MURST) under grant
Cofin98-02-32 and Cofin 98-02-15. Support for CMU and RS was provided
by NASA through grant number GO-06363.01-95A from the Space Telescope
Science Institute, which is operated by AURA, Inc., under NASA
contract NAS~5-26555.

\begin{center}
\begin{tabular}{lllrlll}
\multicolumn{7}{c}{Table 1}\\
\multicolumn{7}{c}{Structural Properties of Low-Redshift Host Galaxies} \\
\hline \\
 Name & z & $\epsilon^{(a)}$ & $C_4^{(b)}$ & $\delta^{(c)}$ & $\Delta^{(d)}$ & $\Delta\theta^{(e)}$ \\
&&&&&& \\ 
 0145+138 & 0.124   &  0.09 & -0.40 &  0.04   &  0.05 &  22.0   \\ 
 0229+200 & 0.139   &  0.16 & -0.50 &  0.02   &  0.03 &  10.0   \\ 
 0347-121 & 0.188   &  0.03 &  0.30 &  0.03   &  0.06 & \nodata \\ 
 0350-371 & 0.165   &  0.25 & -1.20 &  0.02   &  0.02 &   5.0   \\ 
 0521-365 & 0.055   &  0.30 &  0.00 &  0.02   &  0.04 &   4.0   \\ 
 0548-322 & 0.069   &  0.20 & -0.01 &  0.03   &  0.04 &  12.0   \\ 
 0706+591 & 0.125   &  0.15 & -0.80 &  0.02   &  0.04 &  16.0   \\ 
 0806+524 & 0.137   &  0.08 &  1.20 &  0.03   &  0.05 & \nodata \\ 
 0829+046 & 0.180   &  0.10 &  0.20 &  0.04   &  0.04 & \nodata \\ 
 0927+500 & 0.188   &  0.24 &  0.60 &  0.03   &  0.04 &   7.0   \\ 
 1104+384 & 0.031   &  0.19 &  0.00 &  0.03   &  0.04 &  15.0   \\ 
 1136+704 & 0.045   &  0.04 &  0.00 &  0.01   &  0.01 & \nodata \\ 
 1212+078 & 0.136   &  0.04 & -0.10 &  0.01   &  0.01 & \nodata \\ 
 1218+304 & 0.182   &  0.11 & -1.50 &  0.04   &  0.05 &  23.0   \\ 
 1229+643 & 0.164   &  0.16 & -0.30 &  0.02   &  0.03 &  12.0   \\ 
 1255+244 & 0.141   &  0.09 &  0.60 &  0.02   &  0.03 & \nodata \\ 
 1418+546 & 0.152   &  0.30 & -0.30 &  0.02   &  0.02 &  15.0   \\ 
 1426+428 & 0.129   &  0.33 & -0.30 &  0.02   &  0.03 &   3.0   \\ 
 1440+122 & 0.162   &  0.20 &  1.40 &  0.03   &  0.01 &  10.0   \\ 
 1514-241 & 0.049   &  0.02 &  0.20 &  0.02   &  0.03 & \nodata \\ 
 1728+502 & 0.055   &  0.06 &  0.40 &  0.01   &  0.01 & \nodata \\ 
 1807+698 & 0.051   &  0.05 & -0.80 &  0.02   &  0.04 & \nodata \\ 
 1853+671 & 0.212   &  0.15 &  0.40 &  0.05   &  0.04 &  30.0   \\ 
 1959+650 & 0.048   &  0.20 &  1.50 &  0.03   &  0.07 &  17.0   \\ 
 2005-489 & 0.071   &  0.26 &  0.80 &  0.03   &  0.05 &  12.0   \\ 
 2200+420 & 0.069   &  0.25 & -0.90 &  0.02   &  0.05 &   4.0   \\ 
 2201+044 & 0.027   &  0.06 & -0.60 &  0.05   &  0.00 & \nodata \\ 
 2326+174 & 0.213   &  0.12 & -0.90 &  0.04   &  0.03 &  23.0   \\ 
 2344+514 & 0.044   &  0.24 & -0.60 &  0.02   &  0.02 &   5.0   \\ 
 2356-309 & 0.165   &  0.25 &  0.70 &  0.04   &  0.07 &   6.0   \\ 
\hline \\
\multicolumn{7}{l}{$^{(a)}$ Ellipticity of the host galaxy at the effective radius. } \\
\multicolumn{7}{l}{\ \  Typical error $\pm$ 0.02 far from the nucleus.} \\
\multicolumn{7}{l}{$^{(b)}$ Fourier coefficient 100*$C_4$ describing the isophote } \\
\multicolumn{7}{l}{\ \ shape (see text); typical uncertainty is 0.5.}\\
\multicolumn{7}{l}{$^{(c)}$ Relative displacement of isophotes (dimensionless,}\\
\multicolumn{7}{l}{\ \ see definition in \S~3.3); typical uncertainty is $\pm$ 0.01.}\\
\multicolumn{7}{l}{$^{(d)}$ Off-centering (arcsec) of the nucleus with respect to the host  }\\
\multicolumn{7}{l}{\ \  galaxy (see definition in \S~3.3); typical uncertainty is 
0.05 arcsec.}\\
\multicolumn{7}{l}{$^{(e)}$ Maximum observed twisting of isophotes (deg); }\\
\multicolumn{7}{l}{\ \ measured only for host galaxies with $\epsilon >$ 0.1.}\\
\end{tabular} 
\end{center}

\newpage

\begin{center}
{\bf Table 2}\\
\begin{tabular}{llllll}
\multicolumn{6}{c}{Properties of Close Environments of BL Lac Objects} \\
\hline \\
 Name    & Feature            & $m_R$& $M_R^{(a)}$ & $\Delta^{(b)}$ & PA$^{(c)}$ \\
 &  &                           (mag)& (mag) & (\asec/kpc) &(deg) \\
 & & & & \\
0521--365& Optical jet       &  19.9   & \nodata & 1.8  /  2.7 & $\sim$ 305 \\
0706+592 & Compact companion &  24.8   &  -15    & 1.14 /  3.5 & 170 \\
0829+046 & Companion galaxy  &  19.9   &  -20.5  & 4.9  / 20   & 145 \\ 
1229+645 & Companion galaxy  &  19.2   &  -20.9  & 3.4  / 13   & 210 \\  
         & Compact companion &  21.3   &  -18.8  & 4.5  / 17   &   3 \\ 
1426+428 & Compact companion &  21.7   &  -17.9  & 3.9  / 12   &  15 \\ 
1440+122 & Companion  galaxy &  16.7   &  -23.4  & 2.5  /  9.4 & 260 \\ 
         & Compact companion &  19.5   &  -20.6  & 0.3  /  1.1 &  70 \\ 
1959+650 & Dust lane         & \nodata & \nodata & 1.2  /  1.6 & $\sim$ 20 \\ 
1807+698 & Optical jet       &  21.7   & \nodata & 3.1  /  4.2 & $\sim$ 210 \\
1853+671 & Companion  galaxy &  21.8   &  -18.9  & 2.1  /  9.8 & 323 \\
2005--489& Diffuse companion &  22.7   &  -15.5  & 8.5  / 15.6 &  15 \\ 
2201+044 & Optical jet       &  24.2   & \nodata & 2.1  /  1.6 & $\sim$ 315 \\
2326+174 & Compact companion &  23.2   &  -17.5  & 3.2  / 15   & 150 \\ 
2356--309& Compact companion &  22.5   &  -17.3  & 1.2  /  4.6 & 113 \\ 
\hline
\multicolumn{6}{l}{(a) Absolute magnitude assuming the same redshift as the 
BL Lac object.} \\
\multicolumn{6}{l}{(b) Projected distance from the nucleus (in arcsec and kpc) assuming   } \\
\multicolumn{6}{l}{\  assuming the same redshift as the BL Lac object.} \\
\multicolumn{6}{l}{(c) Position angle of the feature.} \\
\end{tabular}	
\end{center}



\begin{figure}
\psfig{file=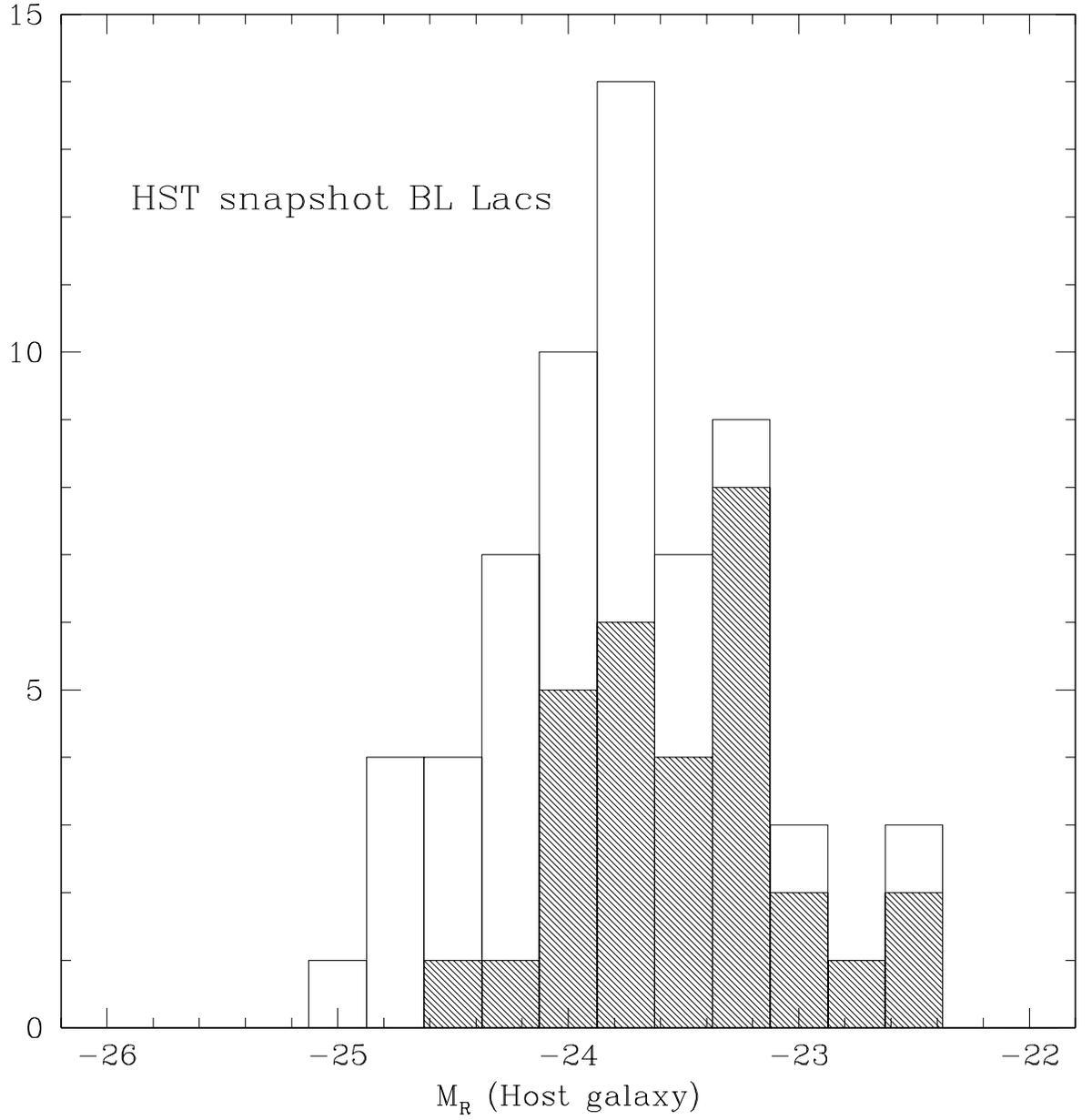}
\caption{Distribution of host galaxy absolute magnitudes
for the full HST snapshot sample (Paper II). 
{\it Hatched area:} The low-redshift subsample discussed in this paper.} 
\end{figure}

\begin{figure}
\psfig{file=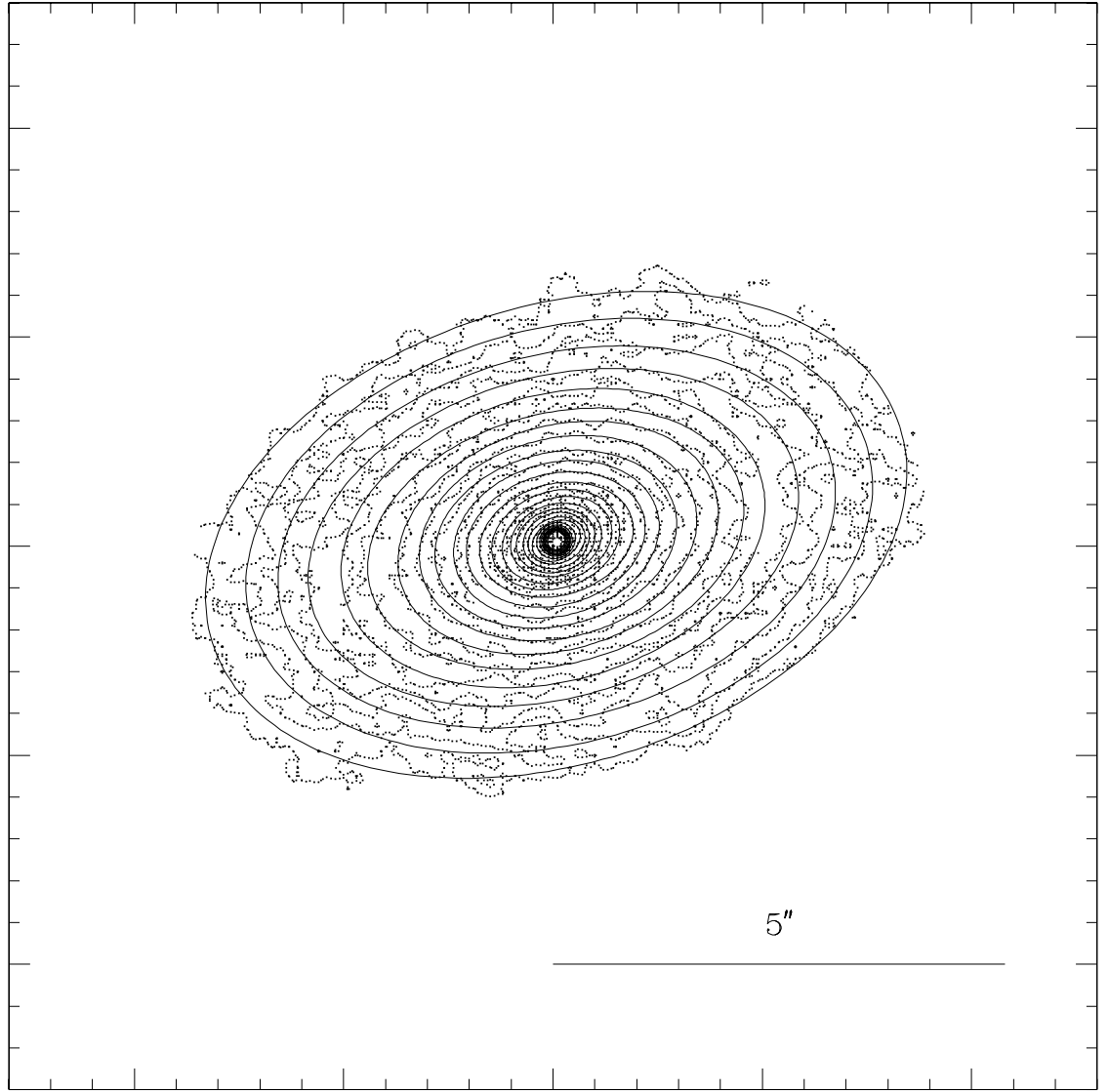}
\caption{Isophotes derived from the surface photometry analysis 
({\it dotted lines}) and fitted ellipses ({\it solid lines}) 
for the BL Lac object 1426+428.
}
\end{figure}

\begin{figure}
\psfig{file=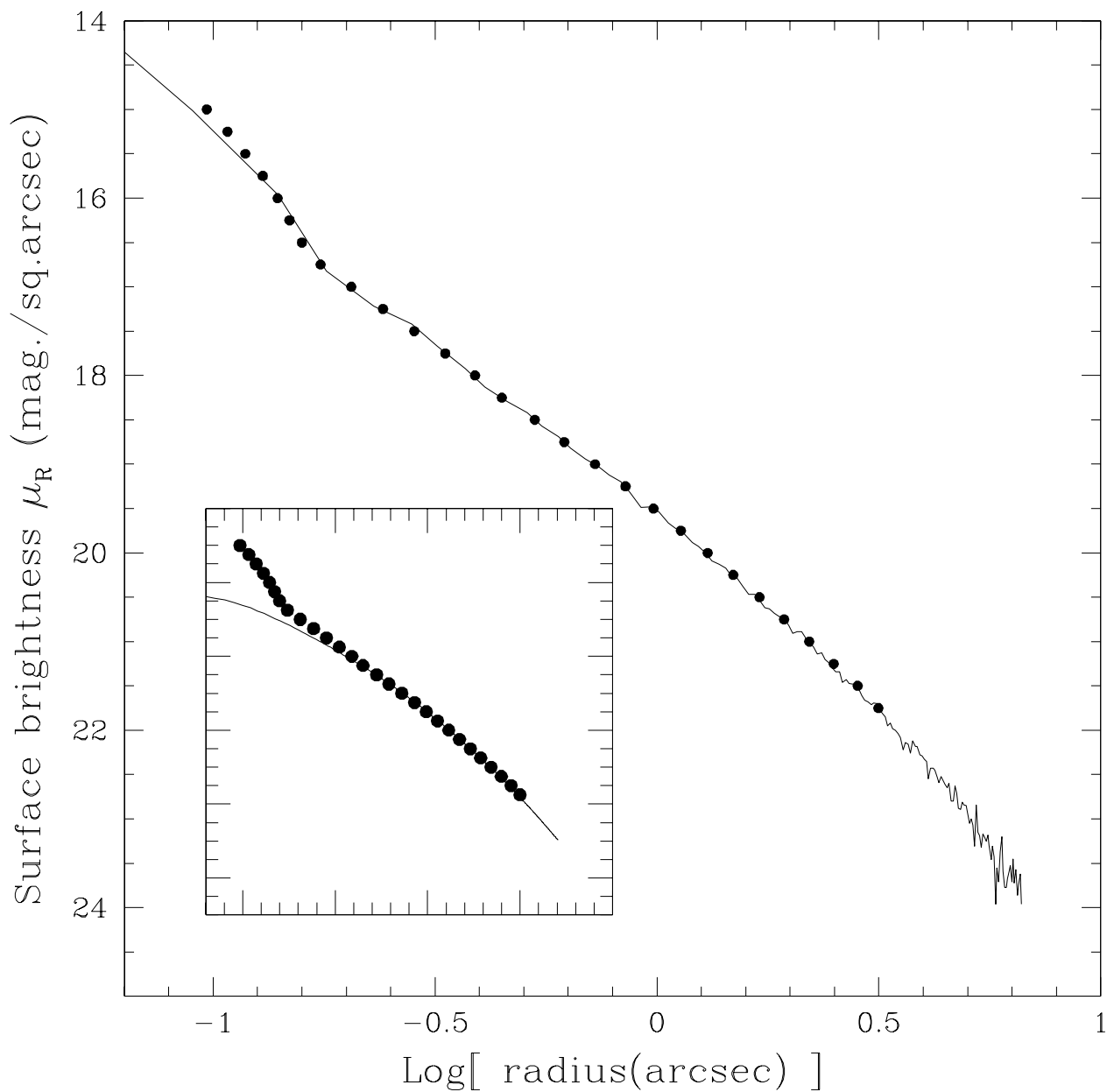}
\caption{The radial surface brightness distribution of the BL Lac 1426+428 
derived from the two-dimensional analysis ({\it filled circles}) 
agrees very well with the azimuthally averaged one-dimensional profile 
({\it solid line}). The inset shows the brightness profile 
({\it filled circles}) compared with the de Vaucouleurs fit ({\it solid line}).
The excess at small radii is due to the presence of the bright nucleus 
(see Paper I for details).
}
\end{figure}

\begin{figure}
\psfig{file=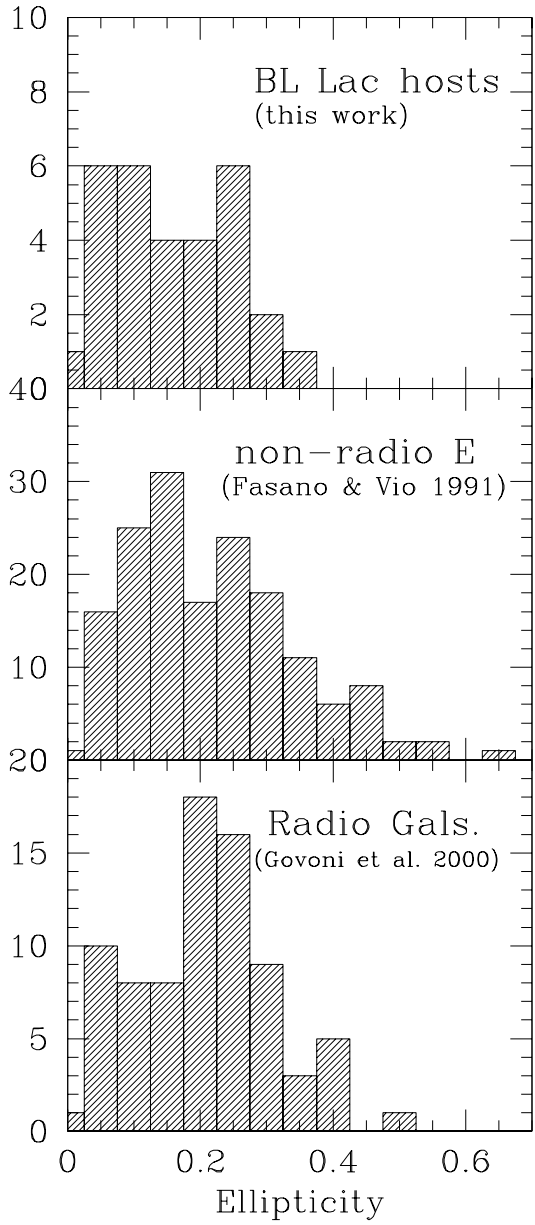}
\caption{Comparison of the distribution of ellipticities for the
host galaxies of BL Lac objects 
({\it top}) with those of normal ellipticals ({\it middle}) 
from a sample of nearby galaxies (Fasano \& Vio 1991) 
and low-redshift radio galaxies 
({\it bottom}; Govoni et al. 2000).
}
\end{figure}

\begin{figure}
\psfig{file=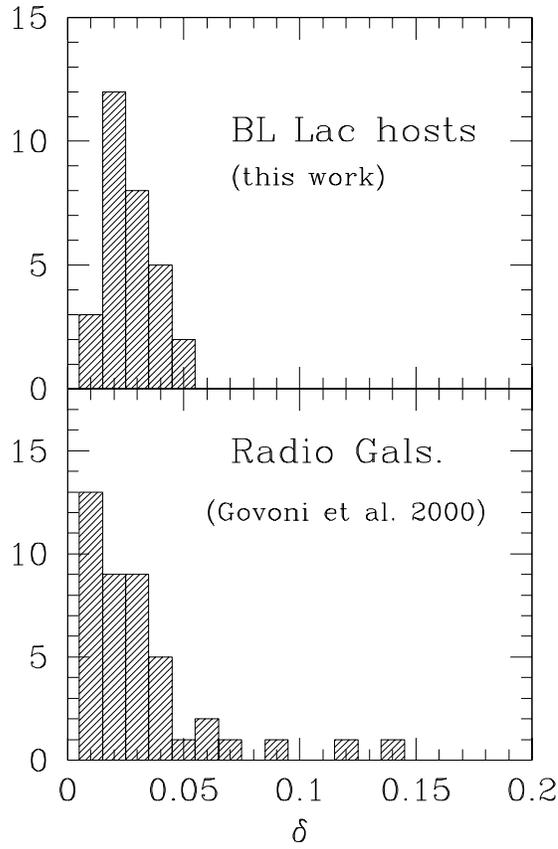}
\caption{Comparison of the isophote displacement, $\delta$ (see text), for 
host galaxies of BL Lacs ({\it top}; present work) and low-redshift 
radio galaxies ({\it bottom}; Govoni et al. 2000).
}
\end{figure}

\begin{figure}
\psfig{file=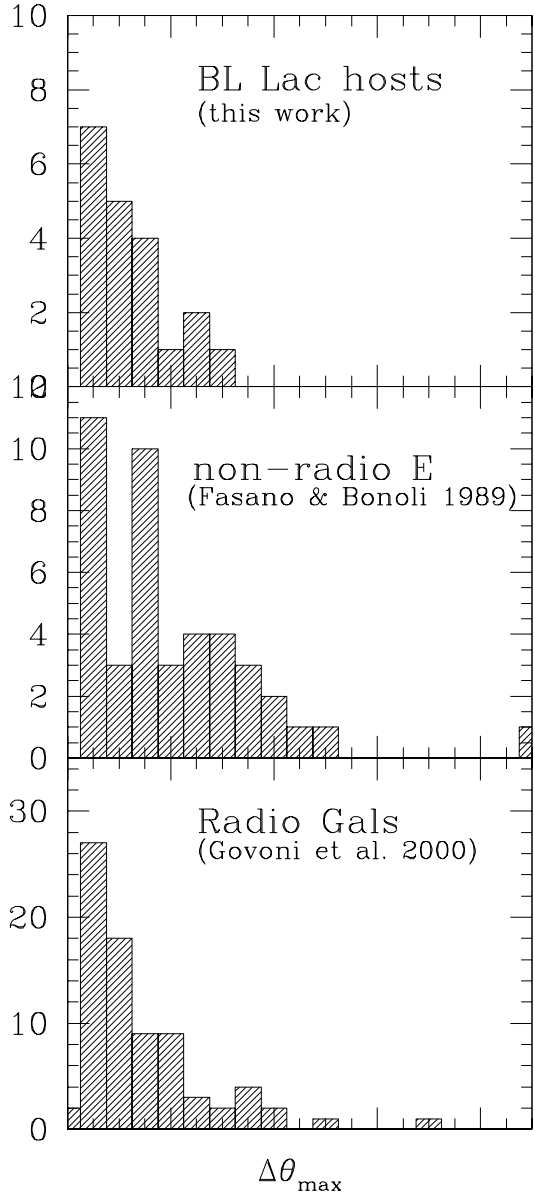}
\caption{Comparison of the maximum isophotal twisting, $\Delta\theta_{max}$,
 for BL Lac host galaxies ({\it top}), 43 non-radio elliptical
 galaxies ({\it middle}; Fasano \& Bonoli 1989), and 
 low-redshift radio galaxies ({\it bottom}; Govoni et al. 2000).}
\end{figure}

\begin{figure}
\psfig{file=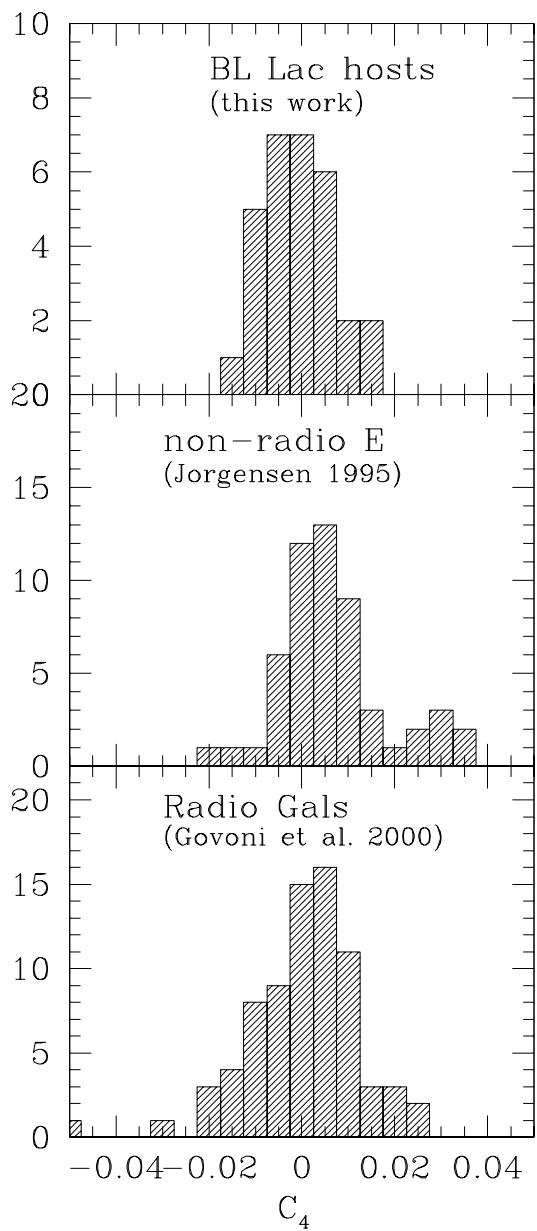}
\caption{Comparison of the $C_4$ parameter (see text) for BL Lac
host galaxies ({\it top}), normal ellipticals and S0 galaxies ({\it middle};
Jorgensen et al. 1995), and low-redshift radio galaxies ({\it bottom}; 
Govoni et al. 2000). The distributions are similar,
and consistent with little or no diskiness or boxiness.}
\end{figure}

\begin{figure}
\caption{Example images of low-redshift HST-observed BL Lac objects 
({\it left panels}) compared to galaxy-model-subtracted images ({\it right
panels}). Both panels are displayed using the same grey-scale.
{\it Top:} PKS~0521-365 and its optical jet and bright knot 
$\sim 2$~arcsec from the nucleus.
{\it Middle:} BL~Lac (2200+420). No significant residuals are 
detected after subtraction of the model (point source plus elliptical
galaxy, convolved with the PSF).
{\it Bottom:} H~2356-309. A compact companion is detected 1.2~arcsec from 
the nucleus of the galaxy. The cross-shaped residuals are
due to the diffraction pattern of the HST PSF, which is not fully described
by the two-dimensional modeling.
Field orientation (arrow mark North) and scale are given for each image.}
\end{figure}

\end{document}